\documentstyle[12pt,axodraw]{article}

\newcommand{\be}{\begin{equation}}
\newcommand{\ee}{\end{equation}}
\newcommand{\bea}{\begin{eqnarray}}
\newcommand{\eea}{\end{eqnarray}}

\newcommand{\h}{\hspace{2mm}}

\newcommand{\unitm}{\hbox{1$\!\!$1}}

\begin{document}
\baselineskip .25in
\newcommand{\numero}{SHEP 95/19}   

\newcommand{\titre}{A dual description of the four dimensional
non-linear sigma model}

\newcommand{\auteura}{Noureddine Mohammedi}
\newcommand{\auteurb}{Richard T. Moss}
\newcommand{\auteurc}{Roger D. Simmons }
\newcommand{\place}{Department of Physics\\University of
Southampton\\ Southampton SO17 1BJ \\ U.K. }
\newcommand{\beq}{\begin{equation}}
\newcommand{\eeq}{\end{equation}}

\newcommand{\abstrait}
{The dual of the four dimensional non-linear sigma model
is constructed using techniques familiar to string theory.
This construction necessitates the introduction of a rank two
antisymmetric tensor field whose properties are examined.
The physics of the dual theory and that of the original
model are compared. As an illustration we study in detail
the SU(2) chiral model. We find that the scattering
amplitudes of the charged Goldstone bosons in the two theories
are in complete agreement at the one loop level.}
\begin{titlepage}
\hfill \numero  \\

\vspace{.5in}
\begin{center}
{\large{\bf \titre }}
\bigskip \\ by \bigskip \\ \auteura \bigskip \\  \auteurb \\
  \bigskip  and \bigskip \\ \auteurc
    \bigskip \\ \place \bigskip \\

\vspace{.9 in}
{\bf Abstract}
\end{center}
\abstrait
 \bigskip \\
\end{titlepage}

\newpage
\section{Introduction}
One of the most striking features of string theory is that it
possesses a dual symmetry which typically reveals itself in two ways.
The first is known as ``T duality'' and is a generic feature of
theories
with compactified spatial dimensions. The simplest example of such
duality manifests itself in the fact that the spectrum obtained
when one
dimension is compactified on a circle of radius $R$ is found to be
indistinguishable from that obtained when the compactification takes
place
on a circle of radius $1/R$ \cite{senda}.
The second type of duality is known as ``S duality''
and interchanges the strong and weak coupling limits of string
theory. This remarkably generalises the strong-weak (or
electric-magnetic)
duality conjuctured many years ago by Olive and Montonen
\cite{olive} and shown
by Osborn \cite{osborn} to hold (if at all) for
$N=4$ supersymmetric gauge theories
only.
\par
Recently, dramatic new evidence for the validity of this conjecture
has emerged from the work of A. Sen \cite{sen} and
furthermore a version of
Olive-Montonen duality was surprisingly found by Seiberg and
Witten in $N=2$ supersymmetric gauge theory in
four dimensions \cite{seiberg}.
\par
These two dualities are, however, discrete symmetries of the
spectrum of
string theory. The question which we would like to ask is can this
duality be implemented at the level of the two dimensional non-linear
sigma model which describes the low energy theory of strings?
Indeed, the dual theory of an arbitrary sigma model with an Abelian
isometry was constructed by Buscher in ref.\cite{buscher}.
There is in fact an
algorithm for constructing the dual theory of a sigma model which
has the advantage of being applicable to sigma models in any
number of dimensions \cite{verlende}.
The algorithm consists of gauging a symmetry (an isometry) of
the action by introducing non-propagating gauge fields and whose field
strength is forced to vanish by means of a Lagrange multiplier - the
original theory being indeed regained if one integrates out the
Lagrange
multiplier. On the other hand, if one integrates over the gauge fields
instead, one obtains the dual theory where the Lagrange multiplier
is now
a dynamical field.
\par
It is this algorithm which we want to apply to a general four
dimensional
sigma model where in this case the Lagrange multiplier is a rank two
antisymmetric field and needs a careful treatment.
\par
Four dimensional sigma models, although non-renormalisable, are of
great
importance in the phenomenology of particle physics on account of the
fact that they describe the
dynamics of pions (and mesons in general) at energies which are small
compared with the inverse confinement radius of QCD.
The Higgs sector
of the Standard Model is another example where four dimensional sigma
models play a crucial role, since in the limit of a very large
Higgs mass,
the Goldstone bosons of the symmetry breaking mechanism may be
described
by a non-linear sigma model as discussed in \cite{gai6,gai11,gai12}.
The connection in this case is made via the equivalence theorem
\cite{gai6,gai5,gai7,gai8} which states that, for a given energy
regime,
the high energy amplitude for a process with external longitudinally
polarised vector bosons is equal to the amplitude of the process in
which the external vector bosons  are replaced by the corresponding
Goldstone bosons. Therefore the study of the dual of the
four dimensional sigma model might have some physical significance
in the
Standard Model.
\par
In this paper we would like to examine the dual theory
of the four dimensional non-linear sigma model. This
study requires a good understanding of the gauge
theory of a rank two antisymmetric tensor field and
we therefore start, in section two, by reviewing this
theory and showing explicitly how the antisymmetric
field could be taken as the dual of a scalar field.
Section three deals with the construction of
the dual theory of a four dimensional sigma model,
which is in fact analogous to the one used in
two dimensional sigma models. We apply this construction
to the $SU(2)$ chiral sigma model in section four where
we calculate the scattering amplitudes at one loop
in the dual theory and compare them with those obtained
from the original theory. Finally we end this paper
with some comments and highlight further issues which
need to be explored.

\section{An alternative description of a scalar field}

As will be made clear shortly, an idea central to our duality
programme will be
the replacement of a scalar field, $\phi$, with a four dimensional
antisymmetric tensor field, $\lambda_{\mu\nu}$. We immediately
have a problem with
the degrees of freedom count - the scalar field has just one,
whilst the
antisymmetric tensor field has six.
It is therefore
very instructive and pedagogical to analyse in some detail
the gauge theory of an antisymmetric tensor field.
Our starting point is the Lagrangian of this theory which
we take to have the form \cite{ramond}
\begin{equation}
{\mathcal L}=-\frac{1}{2}
\epsilon_{\mu\nu\rho\sigma}\partial_{\nu}\lambda_{\rho\sigma}
\epsilon_{\mu\nu'\rho'\sigma'}\partial_{\nu'}\lambda_{\rho'\sigma'}.
\end{equation}
To count the degrees of freedom we can simply
use phase space to identify the true
degrees of freedom, however we prefer to use
a covariant way of doing the counting.
\par
Using the identity
\begin{equation}
\epsilon_{\mu\nu\rho\sigma}\epsilon_{\mu\nu'\rho'\sigma'}=
g_{\nu\nu'}(g_{\rho\rho'}g_{\sigma\sigma'}
-g_{\rho\sigma'}g_{\sigma\rho'})
+g_{\rho\nu'}(g_{\sigma\rho'}g_{\nu\sigma'}
-g_{\sigma\sigma'}g_{\nu\rho'})
+g_{\sigma\nu'}(g_{\nu\rho'}g_{\rho\sigma'}
-g_{\nu\sigma'}g_{\rho\rho'})
\end{equation}
we obtain the following equation of motion for
the antisymmetric field $\lambda_{\rho\sigma}$
\begin{equation}
\Box
\lambda_{\rho\sigma}-
\partial_{\nu}\left(\partial_{\rho}\lambda_{\nu\sigma}
-\partial_{\sigma}\lambda_{\nu\rho}\right)=0.
\end{equation}
These equations (and the action) are left unchanged if we perform
the gauge transformation
\begin{equation}
\lambda_{\rho\sigma}\rightarrow\lambda'_{\rho\sigma}=
\lambda_{\rho\sigma}+\partial_{[\rho}\zeta_{\sigma ]}
\label{gcon0}
\end{equation}
and using this freedom we choose our $\lambda_{\rho\sigma}$ to satsify
(the Lorentz gauge)
\begin{equation}
\partial_{\rho}\lambda_{\rho\sigma}=0.
\label{gcon}
\end{equation}
This has the advantage that it decouples the different components of
$\lambda_{\rho\sigma}$ in a covariant way whilst leaving us with the
simple wave equation
\begin{equation}
\Box\lambda_{\rho\sigma}=0
\end{equation}
which has plane wave solutions of the form\footnote{Although
(for simplicity)
we are using ``unquantised'' language, the extension of the
argument to
the quantised field $\lambda_{\rho\sigma}$ is transparent.}
\begin{equation}
\lambda_{\rho\sigma}={\mathcal V}_{\rho\sigma}\exp(-ik.x)
\end{equation}
where $k^{2}=0$, and ${\mathcal V}_{\rho\sigma}$ contains
the polarisation information of the field. The gauge condition
then leads to the constraints
\begin{equation}
k_{\rho}{\mathcal V}_{\rho\sigma}=0.
\end{equation}
This is a set of four equations of which only three are independent,
hence the polarisation constraints eliminate
three (out of six) degrees of freedom.
To make
the identification with a scalar field we must eliminate still two
more
degrees of freedom and
the required condition comes from the further gauge freedom
allowed by the
constraint equation.  In fact we have not yet exhausted the constraints
imposed by gauge invariance. Within the Lorentz gauge, we are still
free to make another gauge transformation
\begin{equation}
\lambda_{\rho\sigma}\rightarrow
\lambda_{\rho\sigma}+\partial_{[\rho}A_{\sigma]}
\label{gcon2}
\end{equation}
provided we demand that $A_{\rho}$ satisfies
\begin{equation}
\Box A_{\sigma}-\partial_{\rho}\partial_{\sigma}A_{\rho}=0.
\label{max}
\end{equation}
These last equations are simply Maxwell's equations for
electromagnetism
in the vacuum and are themselves left invariant under
\begin{equation}
A_{\rho}\rightarrow A_{\rho}+\partial_{\rho}\psi.
\end{equation}
We are therefore led to gauge fix the gauge fixing conditions
themselves and this will have important consequences when the
theory is quantised - it leads to what are known as ``ghosts for
ghosts''.
\par
As in electromagnetism we can choose the Lorentz gauge for $A_\mu$
such that
\be
\partial_\mu A_\mu=0.
\ee
leading once again to the wave equation,
$\Box A_\mu=0$, and the solution
\begin{equation}
A_{\rho}=\epsilon_{\rho}\exp(-ik.x)
\end{equation}
with $\epsilon_{\rho}$ containing the polarisation information
of $A_\mu$
and again $k^2=0$. The polarisation vector $\epsilon_\mu$ is
constrained
by the gauge choice to satisfy
\be
k_\mu\epsilon_\mu=0.
\ee
This last equation eliminates one component of the gauge function
$A_\mu$.
However, as is well-known in electromagnetism,
the Lorentz gauge does not fix the gauge
transformations completely since we remain in the Lorentz
gauge by performing the gauge transformation
\begin{equation}
A_{\rho}\rightarrow A_{\rho}+\partial_{\rho}\chi
\hspace{1cm}{\mathrm {provided}}\hspace{1cm}
\Box\chi=0.
\end{equation}
The condition  $\Box\chi=0$ also has a wave function
solution $\chi=\exp\left(-ik.x\right)$. This residual gauge
freedom corresponds to changing $\epsilon_\mu$ by a multiple
of $k_\mu$
\be
\epsilon_\mu\rightarrow \epsilon_\mu+\beta k_\mu
\ee
allowing us to kill another component
of the gauge function $A_\mu$, and so, as expected from
electomagnetism, $A_\mu$ has only two independent
polarisations.
\par
Let us now go back to our gauge choice for $\lambda_{\mu\nu}$
and see the effects of all this on the polarisation tensor
${\mathcal V}_{\rho\sigma}$.
For our plane wave solutions, the residual gauge transformation
on the field $\lambda_{\mu\nu}$ amounts to the change
\begin{equation}
{\mathcal V}_{\rho\sigma}\rightarrow {\mathcal V}_{\rho\sigma}+
\left( k_{\rho}\epsilon_{\sigma}-k_{\sigma}\epsilon_{\rho}\right)
\,\,\,\,.
\end{equation}
Since $\epsilon_\mu$ has only two independent components,
this residual gauge transformation can be used to remove
two more degrees of freedom.
Therefore
the overall change has been $6\rightarrow 3\rightarrow 1$ and our
$\lambda_{\rho\sigma}$ now has the correct number of degrees of
freedom to
be seriously taken to represent a real scalar field.
\par
The duality between the antisymmetric tensor field and the
free scalar field can be made more plausible using a path
integral formulation.
Notice that the above action depends only on the ``field strength''
of $\lambda_{\mu\nu}$ namely
$V_\mu = \epsilon_{\mu\nu\rho\sigma}\partial_\nu\lambda_{\rho\sigma}$.
We have simply
\be
{\mathcal L}=-\frac{1}{2}V_\mu V_\mu
\ee
subject to the constraint $\partial_\mu V_\mu=0$.
Including this restriction,
the generating functional for this model is
\bea
{\mathcal Z}&=&
\int {\mathcal D}\lambda
\exp i \left(-\frac{1}{2}
g^2\int{\mathrm d}^4x
\epsilon_{\mu\nu\rho\sigma}\partial_\nu\lambda_{\rho\sigma}
\epsilon_{\mu\nu '\rho '\sigma '}\partial_{\nu '}
\lambda_{\rho '\sigma '}.
\right)\nonumber\\
&=&
\int {\mathcal D}V\delta\left(\partial_\mu V_\mu\right)
\exp i\left(-\frac{1}{2}g^2\int{\mathrm d}^4x
V_\mu V_\mu\right)\,\,\,\,\,.
\label{genny}
\eea
Here $g$ is a dimensionful coupling constant and since the
theory is free, $g$ could be simply absorbed into the
antisymmetric field $\lambda_{\mu\nu}$,
however for later convenience we choose to keep it explicit.
We implement the delta function as usual in the path integral with the
consequent introduction of a
Lagrange multiplier, $\phi$, which can here be interpreted as a
scalar field. We obtain
\begin{eqnarray}
{\mathcal Z}&=&\int {\mathcal D}V{\mathcal D}\phi
\exp i\int{\mathrm d^4}x
\left(-\frac{1}{2}g^2V_\mu V_\mu+\phi\partial_\mu V_\mu\right)
\nonumber\\
&=&\int {\mathcal D}V
\exp-\frac{ ig^2}{2}
\int{\mathrm d^4}x
\left(V_\mu+\frac{1}{g^2}\partial_\mu\phi\right)
\left(V_\mu+\frac{1}{g^2}\partial_\mu\phi\right)
\nonumber\\
&\times &
\int{\mathcal D}\phi\exp \frac{i}{2g^2}\int{\mathrm d^4}x
\left(\partial_\mu\phi\partial^\mu\phi\right).
\end{eqnarray}
The path integral over $V$ is now in the form of a Gaussian
and is just a
number which we normalise to 1 leaving
\be
{\mathcal Z}=\int {\mathcal D}\phi\exp
\frac{i}{2g^2}\partial^\mu\phi\partial_\mu\phi
\ee
which is of the correct form to describe a scalar field.
(We note also that
the scale factor, $g$, has become inverted).

In performing this manipulation we have glossed over an
important point which
we now bring to light\footnote{We thank D. A. Ross for
pointing this out
to us.}.
The generating functional (\ref{genny}) in fact
contains
source terms and (in terms of $\lambda_{\mu\nu}$) we
should have considered
\be
{\mathcal Z}\left[J\right]=\int {\mathcal D}\lambda
\exp i\int{\mathrm d^4}x\left[-\frac{1}{2}
\epsilon_{\mu\nu\rho\sigma}\epsilon_{\mu\nu '\sigma '\rho '}
\partial_\nu
\lambda_{\rho\sigma}\partial_{\nu '}\lambda_{\rho '\sigma '}
+J^{\rho\sigma}\lambda_{\rho\sigma}\right]
\ee
with $J^{\rho\sigma}$ an antisymmetric source. To be able to go from
the path integral integration over $\lambda_{\mu\nu}$ to a
path integral
over $V_\mu$ demands that the source term should be
restricted to have the form
\be
\int {\mathrm d^4}x J^{\rho\sigma}\lambda_{\rho\sigma}=
\int{\mathrm d^4}x
\epsilon^{\mu\nu\rho\sigma}\partial_{\mu}\chi_\nu\lambda_{\rho\sigma}
=-\int{\mathrm d^4}x
\chi_\nu\epsilon^{\mu\nu\rho\sigma}\partial_\mu\lambda_{\rho\sigma}
=\int{\mathrm d^4}x\chi^\mu V_\mu
\ee
to be of a suitable form to act as a source term for the $V_\mu$
 fields.

We therefore see that before our fields can be interpreted as
an alternative
description of scalar fields the source term must have the
very specific
form
\be
J^{\rho\sigma}=\epsilon^{\mu\nu\rho\sigma}\partial_\mu \chi_\nu
\ee
and this causes complications when the antisymmetric tensor field
$\lambda_{\mu\nu}$  is no longer free.
\par
Before leaving this section we would like to examine
the quantisation of the antisymmetric field
$\lambda_{\mu\nu}$ \cite{townsend,siegel}.
We choose to quantise the theory in the
Lorentz gauge $\partial_\mu\lambda_{\mu\nu}=0$
which demands the addition of the gauge fixing term
\be
{\mathcal L}_{{\mathrm g.f}}=
-2\partial_\mu\lambda_{\mu\alpha}\partial_\nu\lambda_{\nu\alpha}
\ee
to our starting action. The ghost Lagrangian corresponding
to this gauge fixing is given by
\be
{\mathcal L}_{{\mathrm gh}}=
C^*_\mu\left(\Box g_{\mu\nu}-\partial_\mu\partial_\nu\right)
C_\nu.
\ee
which is itself invariant
under
\be
\delta C_\mu= \partial_\mu\alpha\,\,\,,\,\,\,
\delta C^*_\mu=\partial_\mu\beta,
\ee
with $\alpha$ and $\beta$ two anticommuting scalar functions.
Since we are integrating over $C_\mu$ and $C^*_\mu$ in the
path integral, we need to fix the ghost gauge freedom, i.e.
we shall be
adding ghosts for ghosts.
We fix the gauge by introducing the term
\be
{\mathcal L}_{{\mathrm gh.g.f}}
=-\partial^\mu C^*_\mu\partial^\nu C_\nu
\ee
and add a corresponding ghost Lagrangian for this new
gauge fixing
\be
{\mathcal L}_{{\mathrm gh.gh}}=
\xi^*\Box\xi + \kappa^*\Box\kappa\,\,\,,
\ee
where $\xi$ and $\kappa$ are
commuting scalars and are the ghosts for ghost fields.
Hence the total effective Lagrangian is given by
\be
{\mathcal L}_{{\mathrm{eff}}}=
\lambda_{\mu\nu}\Box\lambda_{\mu\nu}+
C^*_\mu\Box C_\mu+
\xi^*\Box \xi+
\kappa^*\Box\kappa
\,\,\,\,.
\ee
The quantisation of the free antisymmetric field will be
of use in the following sections.

\section{The Abelian Duality}
In this section we would like to apply the techniques of string theory
in constructing the dual theory of the four dimensional sigma model.
Let us take a target space whose coordinates, $\rho^a$, we split as
$\rho^a=\left(\theta,\phi^i\right)$ and, without loss of generality,
define a four dimensional non-linear sigma model on this space as
\be
S\left(\theta,\phi\right)
=\int {\mathrm{d}}^4x\left[\frac{1}{2}G\left(\phi\right)
\partial_\mu\theta\partial_\mu\theta +
G_i\left(\phi\right)\partial_\mu\phi^i
\partial_\mu\theta
+\frac{1}{2}g_{ij}\left(\phi\right)\partial_\mu
\phi^i\partial_\mu\phi^j\right]\,\,\,.
\label{rupert}
\ee
This action is invariant under the global transformation
$\theta\rightarrow
\theta +\alpha$ and the duality transformation emerges upon minimally
gauging this global symmetry and adding a Lagrange multiplier term
constraining the gauge field to be pure gauge \cite{verlende}.
We therefore consider
\be
S\left(\theta,\phi\right)
=\int {\mathrm{d}}^4x
\left[
\frac{1}{2}G\left(\phi\right)D_\mu\theta D_\mu\theta +
G_i\left(\phi\right)\partial_\mu\phi^iD_\mu\theta +
\frac{1}{2}g_{ij}\left(\phi\right)\partial_\mu
\phi^i\partial_\mu\phi^j -
\frac{1}{2}\epsilon_{\mu\nu\rho\sigma}\lambda_{\rho\sigma}F_{\mu\nu}
\right]
\label{noddy}
\ee
where $D_\mu\theta=\partial_\mu\theta+A_\mu$ and $F_{\mu\nu}=
\partial_\mu A_\nu-\partial_\nu A_\mu$.
Variation with respect
to $\lambda_{\rho\sigma}$ imposes the constaint $F_{\mu\nu}=0$
which is solved by
$A_\mu=\partial_\mu\xi$ which in turn gives
\be
D_\mu\theta=\partial_\mu\theta+A_\mu\h =\h\partial_\mu(\theta+\xi)
\ee
i.e. the effect of rewriting the model in a gauged form has been to
replace $\theta$ by $\theta +\xi$ which is dynamically of
no consequence. Hence eliminating the Lagrange multiplier
takes us back to the original theory.
\par
Next, keeping the Lagrange multiplier and varying instead
with respect to the gauge field gives,
after integration by parts,
\be
\frac{\delta{\mathcal L}}{\delta A_\mu} = G\left(\partial_\mu\theta
+ A_\mu\right) +G_i\partial_\mu\phi^i
- \epsilon_{\mu\nu\rho\sigma}\partial_\nu\lambda_{\rho\sigma}=0
\ee
or
\be
D_\mu\theta=\frac{1}{G}\left(
\epsilon_{\mu\nu\rho\sigma}\partial_\nu\lambda_{\rho\sigma}-
G_i\partial_\mu\phi^i
\right).
\ee
Substituting this back into (\ref{noddy}) we obtain our form for
the dual action
\be
S\left(\lambda,\phi\right)= \int{\mathrm d}^4x
\left[ \frac{1}{2}
G_{ij}
\partial_\mu\phi^i\partial^\mu\phi^j
-\frac{1}{2G}\epsilon_{\mu\nu\rho\sigma}\partial_\nu
\lambda_{\rho\sigma}
\epsilon_{\mu\nu '\rho '\sigma '}\partial_{\nu '}
\lambda_{\rho '\sigma '}
+\frac{1}{G}
\epsilon_{\mu\nu\rho\sigma}\partial_\nu\lambda_{\rho\sigma}
G_i\partial_\mu\phi^i
\right]
\label{lumpy}
\ee
where $G_{ij}$ is given by
\be
G_{ij}=g_{ij}-\frac{1}{G}G_iG_j.
\label{stumpy}
\ee
Therefore the dual action describes a non-linear sigma model with
metric $G_{ij}$ interacting with a dynamical antisymmetric
tensor field. It could be thought that the antisymmetric field
$\lambda_{\mu\nu}$ has replaced the field $\theta$ in the original
model.
We also notice that the coupling $G$ has become inverted in various
terms and take this to imply that certain strong coupling behaviour
will now be described correctly with techniques appropriate to small
coupling as expected from duality.
\par
The duality in the action $S\left(\lambda,\phi\right)$ can be seen from
the fact that this action remains invariant under the interchange
\be
\epsilon_{\mu\nu\rho\sigma}\partial_\nu\lambda_{\rho\sigma}\,\,\,
\leftrightarrow\,\,\,G_i\partial_\mu\phi^i,
\ee
which is in the spirit of the duality encountered
in electomagnetism between the electric and magnetic fields.
\par
Another appealing feature of the dual action
$S\left(\lambda,\phi\right)$
can be seen by casting it in the form
\be
S\left(\lambda,\phi\right)= \int{\mathrm d}^4x
\left[ \frac{1}{2}
G_{ij}
\partial_\mu\phi^i\partial^\mu\phi^j
-\frac{1}{2G}\epsilon_{\mu\nu\rho\sigma}\partial_\nu
\lambda_{\rho\sigma}
\epsilon_{\mu\nu '\rho '\sigma '}\partial_{\nu '}\lambda_{\rho'\sigma'}
+\lambda_{\mu\nu}\Omega_{\mu\nu}\right]
\ee
where $\Omega_{\mu\nu}=\epsilon_{\mu\nu\rho\sigma}\partial_
\rho\left(\frac{1}{G}G_i\partial_\sigma\phi\right)$. In this form
the model describes a hydrodynamical flow in the presence of
a vortex $\Omega_{\mu\nu}$; the antisymmetric field $\lambda_{\mu\nu}$
is then the velocity potential and $V_{\mu}$ is the velocity vector
satisfying the continuity equation $\partial_\mu V_\mu=0$.
A similar phenomenon has been shown to exist in
the Abelian Higgs model by Sugamoto \cite{sugamoto}.

\section{Application to SU(2)}

We turn now to investigate some phenomenological
implications of what we
have done. For definiteness we restrict our analysis to the
$SU(2)$ case although the generalisation to other Lie algebras
is staightforward.
We take as our starting point the non-linear sigma model
parameterised by the matrix field $U(x)$ belonging to the
quotient space
$SU(2)_L\otimes SU(2)_R / SU(2)_{L+R}$,
\be
U(x)=\exp\left( i\tau^a\xi_a/\Lambda\right),
\label{oldpar}
\ee
where $\xi^a$, $a=1,2,3$, are the Goldstone bosons associated with the
symmetry breaking $SU(2)_L\otimes SU(2)_R\rightarrow SU(2)_{L+R}$,
$\tau^a$
are the $2\times 2$ Pauli matrices and $\Lambda$ some energy scale.
The current
interest in such models stems from the fact that they can be used to
investigate the symmetry breaking sector of the Standard Model - the
connection being made via the equivalence theorem which relates the
scattering of longitudinally polarised weak vector bosons to those
involving the Goldstone bosons associated with the above symmetry
breaking
pattern - these points are lucidly discussed in \cite{dobado}.

Under an $SU(2)_L\otimes SU(2)_R$ transformation the matrix $U$
transforms
as $U\rightarrow LUR^\dagger$ and the $SU(2)$ invariant sigma
model can
be written
\be
{\mathcal L}=\frac{\Lambda^2}{4}{\mathrm Tr}\partial_\mu U
\partial^\mu U^{-1}
\label{oldfriend}
\ee
where we let $\Lambda$ take any value. (When applied to the symmetry
breaking sector of the Standard Model $\Lambda$ becomes fixed
at the Higgs
VEV scale of 246 GeV). Our interest is in the dual version
of the model and
in order to get this we must first massage (\ref{oldfriend})
into the form
(\ref{rupert}) which we
achieve by separating out one field - we choose that associated with
$\tau^3$ and write
\be
U(x)=
\exp\left(i\theta\tau_3/\Lambda\right)
\exp\left(i\tau^j\phi_j/\Lambda\right)
\label{newpar}
\ee
where $j$ runs over the values 1 and 2 only. This
new parametrisation can be thought of as
a change of variables from $\xi^a$ to $\theta$ and $\phi^i$.

Using the closed form expression for the exponential
of any linear combination of Pauli matrices
\be
\exp\left(i\tau^\alpha\lambda_\alpha/\Lambda\right)=
\unitm\cos\Omega+i\tau ^\alpha\lambda_\alpha
\frac{1}{\Lambda}
\frac{\sin\Omega}{\Omega}\h\h\h {\mathrm {with}}\h\h\h
\Omega^2=\frac{\lambda^\alpha\lambda_\alpha}{\Lambda^2}
\ee
we arrive at the following form of the $SU(2)$ non-linear sigma model
\begin{eqnarray}
{\mathcal L}&=&
\frac{1}{2}\partial^\mu\theta\partial_\mu\theta
+\frac{1}{2}\frac{1}{\Lambda}\frac{\sin^2\Omega}{\Omega^2}
\epsilon_{ij}\pi^j\partial^\mu\theta\partial_\mu\pi^i
+\frac{1}{4}\frac{\sin^2\Omega}{\Omega^2} \delta_{ij} \partial^\mu\pi^i
\partial_\mu\pi^j
\nonumber\\
&+&\frac{1}{8\Lambda^2}\frac{1}{\Omega^2}\left[
1-\frac{\sin^2\Omega}{\Omega^2}\right]\delta_{ij}\delta_{kl}
\pi^i\pi^k
\partial^\mu\pi^j\partial_\mu\pi^l
\label{splurge}
\end{eqnarray}
where we have defined $\pi^{\pm}=
\left(\phi_1\pm i \phi_2\right)$ and $\Omega^2=\pi^+\pi^-/\Lambda^2$.
The target space indices are contacted by the delta function,
$\delta_{+-}=1$,
and $\epsilon_{+-}=-\epsilon_{-+}=i$.
This action can be seen to be in the form of (\ref{rupert}) with
the identifications
\begin{eqnarray}
g_{ij}&=&\frac{1}{2}
\frac{\sin^2\Omega}{\Omega^2}\delta_{ij}+
\frac{1}{4}\frac{1}{\Omega^2\Lambda^2}\left[
1-\frac{\sin^2\Omega}{\Omega^2}\right]
\delta_{ik}\delta_{jl}\pi^k\pi^l
\nonumber\\
G_{i}&=&\frac{1}{2}\frac{1}{\Lambda}\epsilon_{ij}
\pi^{j}\frac{\sin^2\Omega}{\Omega^2}
\,\,\,,\,\,\, G=1\,\,\,\,.
\label{metricelements}
\end{eqnarray}
It is well known that in two dimensions this model is
renormalisable in the
sense that the counterterms can be absorbed into the terms
already present
in the tree level Lagrangian. In four dimensions, however, this is no
longer true. To get around this problem we take (\ref{splurge}) to
be the first term
in a general momentum expansion with an infinte number of terms and an
infinite number of arbitrary parameters.
The addition of these terms is necessary in order to absorb the
higher dimensional divergences which one encounters in four
dimensional sigma models.
Now when we calculate divergent
quantities we can renormalise the higher order coefficients
thereby making
the theory finite - i.e. the higher order terms are demanded
if the theory
is to make sense.
\par
Our aim is to calculate two-particle scattering amplitudes of the sigma
model in the parametrisation (\ref{newpar})
and to make a comparison with the results obtained using
the dual theory. For this we
enlarge (\ref{splurge}) to include $O(p^4)$ terms and
expand the principal sigma model up to four point interactions.
We have then the Lagrangian obtained by expanding the principal
chiral sigma model up to four fields
\begin{equation}
{\mathcal L}=
	+\frac{1}{2}\partial_{\mu}\theta\partial^{\mu}\theta
	+\frac{1}{4}\delta_{ij}\partial_{\mu}\pi^i\partial^{\mu}\pi^j
	+\frac{1}{2}\frac{1}{\Lambda}\epsilon_{ij}\pi^j
	\partial_{\mu}\theta\partial^{\mu}\pi^i
	-\frac{1}{24\Lambda^2}
	\left[
	\delta_{ij}\delta_{lm}-\delta_{il}\delta_{jm}
	\right]
	\pi^i\pi^j\partial_{\mu}\pi^l\partial^{\mu}\pi^m
\end{equation}
plus the counterterm Lagrangian
\bea
{\mathcal L}_{\mathrm{c.t.}}&\ &=
\frac{4}{\Lambda^4}(M+N)\partial_{\mu}
	\theta\partial^{\mu}\theta
	\partial_{\nu}\theta\partial^{\nu}\theta
	+\frac{8}{\Lambda^4}\delta_{ij}
	\left[
	M\partial_{\mu}\theta\partial^{\mu}\theta
	 \partial_{\nu}\pi^i\partial^{\nu}\pi^j
	+N\partial_{\mu}\theta\partial^{\mu}\pi^i
	  \partial_{\nu}\theta\partial^{\nu}\pi^j
	\right] \nonumber\\
	&\ &+\frac{4}{\Lambda^4}
	\left[
	M\delta_{ij}\delta_{lm}
	+N\delta_{il}\delta_{jm}
	\right]
	\partial_{\mu}\pi^i\partial^{\mu}\pi^j
	\partial_{\nu}\pi^l\partial^{\nu}\pi^m
\label{bigmess}
\eea
where $M$ and $N$ are the arbitrary coefficients of
the $O(p^4)$ contributions.
We stop at just four point terms since the processes of greatest
phenomenological interest are the two-particle scattering
amplitudes on account
of the fact that the equivalence theorem directly relates
these amplitudes
to the amplitudes for scattering processes of the form
$W^i_LW^j_L\rightarrow W^m_LW^n_L$ where
$W^i=W^\pm$, $Z^0$.
To one loop we need to consider 24 Feynman diagrams
(compared with just three
in the more conventional parametrisation (\ref{oldpar}))
a representative selection of which are given in fig.1.
We obtain
\begin{eqnarray}
{\mathcal M}\left(\pi^+\pi^-\rightarrow\pi^+\pi^-\right)&=&
	-i\frac{u}{\Lambda^2}+\frac{4i}{\Lambda^4}
	\left[
	2M_R(s^2+t^2)+N_R(s^2+t^2+2u^2)
	\right]\nonumber\\
	\ & \ &-\frac{i}{(4\pi)^2\Lambda^4}
	\left(
	\frac{1}{12}(9s^2+u^2-t^2)\ln\frac{-s}{\mu^2}\right.
\nonumber\\
	\ & \ &\left.+\frac{1}{12}(9t^2+u^2-s^2)\ln\frac{-t}{\mu^2}
	+\frac{1}{2}u^2\ln\frac{-u}{\mu^2}
	\right)\nonumber\\
\ & \ &\nonumber\\
\ & \ &\nonumber\\
{\mathcal M}\left(\pi^+\pi^-\rightarrow\theta\theta\right) &=&
	i\frac{s}{\Lambda^2}+\frac{4i}{\Lambda^4}
	\left[
	2M_Rs^2+N_R(t^2+u^2)
	\right]\nonumber\\
	\ & \ &-\frac{i}{(4\pi)^2\Lambda^4}
	\left(
	\frac{1}{12}(3t^2+u^2-s^2)\ln\frac{-t}{\mu^2}
	\right. \nonumber\\
	\ & \ & +\left.\frac{1}{12}(3u^2+t^2-s^2)\ln\frac{-u}{\mu^2}
	+\frac{1}{2}s^2\ln\frac{-s}{\mu^2}
	\right)\nonumber\\
\ &\ &\nonumber\\
\ & \ &\nonumber\\
{\mathcal M}\left(\theta\theta\rightarrow\theta\theta\right)
&=&\frac{8i}{\Lambda^4}
	\left[M_R+N_R \right] \left(s^2+t^2+u^2 \right)\nonumber\\
	\ & \ &+\frac{i}{(4\pi)^2\Lambda^4)}
	\left(
	-s^2\ln\frac{-s}{\mu^2}
	-t^2\ln\frac{-t}{\mu^2}
	-u^2\ln\frac{-u}{\mu^2}
	\right).
\end{eqnarray}
Here $s$, $t$ and $u$ are the usual Mandelstam variables, $\mu$ is an
arbitrary renormalisation scale and using dimensional regularisation
in $d=4-2\epsilon$ dimensions we have
the renormalised forms of $M$ and $N$
(using modified minimal subtraction, $\overline{MS}$):
\begin{eqnarray}
M_R&=&M+\frac{1}{24}\frac{1}{(4\pi)^2}\left(
\frac{1}{\epsilon}-\gamma+\ln (4\pi)\right)
\nonumber\\
N_R&=&N+\frac{1}{12}\frac{1}{(4\pi)^2}\left(
\frac{1}{\epsilon}-\gamma+\ln (4\pi)\right).
\label{renpara}
\end{eqnarray}
These scattering amplitudes are in precise agreement with
the results presented in \cite{dobado}.

Let us now turn to the calculation of the equivalent
scattering amplitudes in the dual model. Using the form of
the metric elements given in (\ref{metricelements})
and using (\ref{lumpy})
we can move directly to
the dual description and consider the Lagrangian
\begin{eqnarray}
{\mathcal L}^D&=&
-\frac{1}{2}\epsilon_{\mu\nu\rho\sigma}\partial_\nu\lambda_{\rho\sigma}
\epsilon_{\mu \nu '\rho '\sigma '}\partial_{\nu '}
\lambda_{\rho '\sigma '}
+\frac{1}{2}
\frac{1}{\Lambda}\epsilon_{\mu\nu\rho\sigma}\partial_\nu
\lambda_{\rho\sigma}
\epsilon_{ij}\pi^j\partial^\mu\pi^i\frac{\sin^2\Omega}{\Omega^2}
\nonumber\\
&+&\frac{1}{4}\partial^\mu\pi .\partial_\mu\pi\frac{\sin^2\Omega}
{\Omega^2}
\left(1-\sin^2\Omega\right)
+\frac{1}{8}\frac{1}{\Lambda^2\Omega^2}\left[
1-\frac{\sin^2\Omega}{\Omega^2}+\frac{\sin^4\Omega}{\Omega^2}\right]
\left( \pi.\partial_\mu\pi\right)^2.
\end{eqnarray}
Expanding to four point terms we have
\begin{eqnarray}
{\mathcal L}&=&
-\frac{1}{2}\epsilon_{\mu\nu\rho\sigma}\partial_\nu
\lambda_{\rho\sigma}
\epsilon_{\mu\nu '\rho '\sigma '}\partial_{\nu '}
\lambda_{\rho '\sigma '}
+\frac{1}{2}\frac{1}{\Lambda}\epsilon_{\mu\nu\rho\sigma}
\epsilon_{ij}\pi^j
\partial_\nu\lambda_{\rho\sigma}\partial^\mu\pi^i\nonumber\\
&+&\frac{1}{4}\partial^\mu\pi^i\partial_\mu\pi^i
+\frac{1}{6}\frac{1}{\Lambda^2}
\left(\delta_{ik}\delta_{jl}-\delta_{ij}\delta_{kl}\right)
\pi^i\pi^j\partial^\mu\pi^k\partial_\mu\pi^l
\end{eqnarray}
giving at tree level the independent matrix elements
\[
{\mathcal M}\left(\pi^+\pi^-\rightarrow\pi^+\pi^-\right)
=-i\frac{u}{\Lambda^2},\h\h\h
{\mathcal M}\left(\pi^+\pi^-\rightarrow\lambda_{\mu\nu}
\lambda_{\rho\sigma}\right)
=i\frac{s}{\Lambda^2}
\]
\be
{\mathcal M}\left(\lambda_{\mu\nu}\lambda_{\rho\sigma}
\rightarrow\lambda_{\mu'\nu'}\lambda_{\rho'\sigma'}\right)=0
\ee
in precise agreement with the original model. To get this we have used
the sum over polarisation vectors, ${\mathcal V}_{\rho\sigma}$,
for  external
$\lambda_{\rho\sigma}$ fields
\be
\sum_\varepsilon {\mathcal V}_{\rho\sigma}(\varepsilon){
\mathcal V}^*_{\rho '\sigma '}(\varepsilon)
=\frac{1}{4}\left(
g_{\rho\rho '}g_{\sigma\sigma '}-g_{\rho\sigma '}g_{\sigma\rho '}
\right)
\ee
where $\varepsilon$ labels the helicity of the lines. At tree level,
therefore, we are allowed to make the identification that the
one physical
degree of freedom associated with $\lambda_{\rho\sigma}$ is
the Goldstone
boson $Z^0$.

Proceeding to the higher order corrections we have calculated
the scattering amplitudes for the charged pions in the dual model.
The Feynman diagrams which need to be considered in this case are
given in fig.2 {\it and} fig.3, where we note in particular the
non-zero
contribution of the one-particle reducible diagrams.
The net result due to these diagrams is given by
\begin{eqnarray}
{\mathcal M}\left(\pi^+\pi^-\rightarrow\pi^+\pi^-\right)&=&
	-i\frac{u}{\Lambda^2}+\frac{4i}{\Lambda^4}
	\left[
	2M_R(s^2+t^2)+N_R(s^2+t^2+2u^2)
	\right]\nonumber\\
	\ & \ &-\frac{i}{(4\pi)^2\Lambda^4}
	\left(
	\frac{1}{12}(9s^2+u^2-t^2)\ln\frac{-s}{\mu^2}\right.
\nonumber\\
	\ & \ &\left.+\frac{1}{12}(9t^2+u^2-s^2)\ln\frac{-t}{\mu^2}
	+\frac{1}{2}u^2\ln\frac{-u}{\mu^2}
	\right)\nonumber\\
\ & \ &\nonumber\\
\ & \ &\nonumber\\
{\mathcal M}\left(\pi^+\pi^-\rightarrow\lambda\lambda\right) &=&
	i\frac{s}{\Lambda^2}+\frac{4i}{\Lambda^4}
	\left[
	2M_Rs^2+N_R(t^2+u^2)
	\right]\nonumber\\
	\ & \ &-\frac{i}{(4\pi)^2\Lambda^4}
	\left(
	\frac{1}{12}(3t^2+u^2-s^2)\ln\frac{-t}{\mu^2}
	\right. \nonumber\\
	\ & \ & +\left.\frac{1}{12}(3u^2+t^2-s^2)\ln\frac{-u}{\mu^2}
	+\frac{1}{2}s^2\ln\frac{-s}{\mu^2}
	\right)\nonumber\\
\end{eqnarray}
and again these expressions are precisely those found in the original
model.
Hence, to one loop, the scattering amplitudes of the charged pions
in both model do agree. The scattering amplitude for four external
antisymmetric tensor fields is very much more involved and
will be treated
elsewhere.
\par
Notice that these scattering amplitudes involve the renormalised
parameters $M_R$ and $N_R$ as given in (\ref{renpara}). This means that
we have actually added counterterms of dimension four to the
dual theory.
These terms are found by taking (\ref{bigmess}) and replacing
$\partial_\mu\theta$ by $\epsilon_{\mu\nu\rho\sigma}\partial_\nu
\lambda_{\rho\sigma}$.
\section{Conclusions}
We have studied, at the quantum level, the Abelian gauge theory
of a rank two antisymmetric tensor field non-trivially interacting
with scalar fields in the form of a non-linear sigma model.
This theory is the dual theory of a four
dimensional sigma model obtained using techniques of
two dimensional theories and we have shown that the scattering
amplitudes of the charged pions are the same, at the one loop
level, in the dual and the original theories.
\par
It is therefore clear that the duality transformations  of the
four dimensional sigma model do not change the physics of the
original theory. This is in contrast to the two dimensional case
where the geometry and the physics of the dual theory are
completely changed. This change is in fact due to the
presence of the Wess-Zumino-Witten term in two dimensions.
The dual two dimensional sigma model would be
trivial if one set the Wess-Zumino-Witten term to zero.
The question we would like to address now is could the
inclusion of the Wess-Zumino-Witten term in four dimensions
be of any consequence to the physics of the pions?
\par
If chiral Lagrangians are to be taken as effective theories
of QCD then they ought to incorporate all relevant symmetries
of QCD and the presence of a Wess-Zumino-Witten term is then
essential for the preservation of the symmetries
of QCD \cite{witten}. In four dimensions, and in the
notation of equation (\ref{rupert}), this term can be written
in the form
\be
S_{\mathrm{wzw}}\left(\theta,\phi\right)=
\int{\mathrm{d}}^4x\epsilon_{\mu\nu\rho\sigma}
\left[b_{ijk}\left(\phi\right)
\partial_\mu\theta\partial_\nu\phi^i\partial_\rho\phi^j
\partial_\rho\phi^k +
B_{ijkl}\left(\phi\right)
\partial_\mu\phi^i\partial_\nu\phi^j
\partial_\rho\phi^k \partial_\sigma\phi^l\right]
\ee
with $b_{ijk}$ and $B_{ijkl}$ totally antisymmetric tensors. (This
demand for antisymmetry on the tensors makes it clear that for the
case of $SU(2)$ no non-zero Wess-Zumino-Witten term can be generated).
Adding this term to the sigma model action in
(\ref{rupert}) and performing the
duality transformation leads to the action
\be
I\left(\lambda,\phi\right)=
\widetilde S\left(\theta,\phi\right)
+\int {\mathrm{d}}^4x\epsilon_{\mu\nu\rho\sigma}
B_{ijkl}\partial_\mu\phi^i\partial_\nu\phi^j
\partial_\rho\phi^k \partial_\sigma\phi^l\,,
\ee
where $\widetilde S\left(\theta,\phi\right)$
is obtained from $S\left(\lambda,\phi\right)$ in
(\ref{noddy}) upon making the substitution
\be
G_i\partial_\mu\phi^i\,\,\,\,
\rightarrow\,\,\,\,G_i\partial_\mu\phi^i +
\epsilon_{\mu\nu\rho\sigma}
b_{ijk}\partial_\nu\phi^i\partial_\rho\phi^j
\partial_\rho\phi^k \,.
\ee
Notice that terms of dimension four and six are generated
in the dual theory - we certainly expect
these higher dimensional terms to contribute  beyond the
one loop level.
This issue is currently under investigation.
\par
Another problem which is at the heart of dual theories
is the infrared behaviour of these theories. It was
shown in \cite{rajeev} that the inclusion of the Wess-Zumino-Witten
term in the four dimensional sigma model leads to a non-trivial
infrared fixed point, in addition to the usual Gaussian fixed point.
This is a phenomenon that was shown to occur at three loops by
analytically continuing the theory  to a dimension less than four.
We expect that the terms in the dual theory which are of dimension
six will have dramatic consequences on the infrared behaviour
of the theory at three loops.

\begin{center}
{\bf{Acknowledgements}}
\end{center}
We gratefully acknowledge many productive
conversations with Ken Barnes, Tim Morris and Douglas
Ross, and financial support for RTM and RDS from PPARC.

\newpage
\section*{Representative one-loop diagrams referred to in the text}
\begin{center}
\begin{picture}(400,100)(0,0)

\Curve{(30,50)(50,70)(70,50)}
\Curve{(30,50)(50,30)(70,50)}
\ArrowLine(0,80)(30,50)
\ArrowLine(0,20)(30,50)
\ArrowLine(100,80)(70,50)
\ArrowLine(100,20)(70,50)
\Vertex(30,50){2}
\Vertex(70,50){2}

\ArrowLine(160,90)(180,70)
\Line(180,70)(200,50)
\ArrowLine(160,10)(180,30)
\Line(180,30)(200,50)
\ArrowLine(240,90)(200,50)
\ArrowLine(240,10)(200,50)
\Vertex(200,50){2}
\DashArrowLine(180,70)(180,30){3}
\Vertex(180,70){2}
\Vertex(180,30){2}

\ArrowLine(300,80)(333,80)
\DashArrowLine(400,80)(366,80){3}
\ArrowLine(366,80)(333,80)
\ArrowLine(300,20)(333,20)
\DashArrowLine(400,20)(366,20){3}
\ArrowLine(366,20)(333,20)
\DashArrowLine(333,80)(333,20){3}
\ArrowLine(366,80)(366,20)
\Vertex(333,80){2}
\Vertex(333,20){2}
\Vertex(366,80){2}
\Vertex(366,20){2}
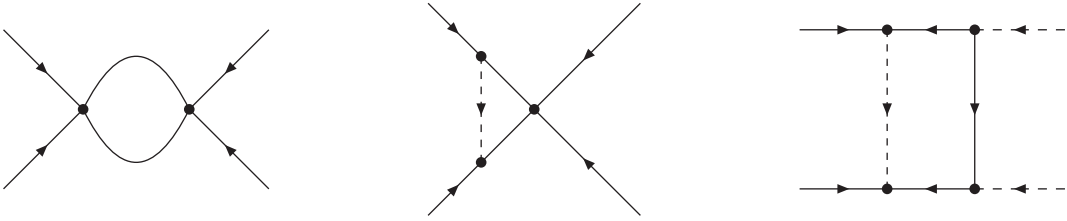
\end{picture}\\ {\sl fig.1}
\end{center}

\begin{center}
\begin{picture}(400,100)(0,0)

\Curve{(30,50)(50,70)(70,50)}
\Curve{(30,50)(50,30)(70,50)}
\ArrowLine(0,80)(30,50)
\ArrowLine(0,20)(30,50)
\ArrowLine(100,80)(70,50)
\ArrowLine(100,20)(70,50)
\Vertex(30,50){2}
\Vertex(70,50){2}

\ArrowLine(160,90)(180,70)
\Line(180,70)(200,50)
\ArrowLine(160,10)(180,30)
\Line(180,30)(200,50)
\ArrowLine(240,90)(200,50)
\ArrowLine(240,10)(200,50)
\Vertex(200,50){2}
\Gluon(180,30)(180,70){4}{4}
\Vertex(180,70){2}
\Vertex(180,30){2}

\ArrowLine(300,80)(333,80)
\ArrowLine(400,80)(366,80)
\ArrowLine(366,80)(333,80)
\ArrowLine(300,20)(333,20)
\ArrowLine(400,20)(366,20)
\ArrowLine(366,20)(333,20)
\Gluon(333,80)(333,20){5}{5}
\Gluon(366,80)(366,20){5}{5}
\Vertex(333,80){2}
\Vertex(333,20){2}
\Vertex(366,80){2}
\Vertex(366,20){2}
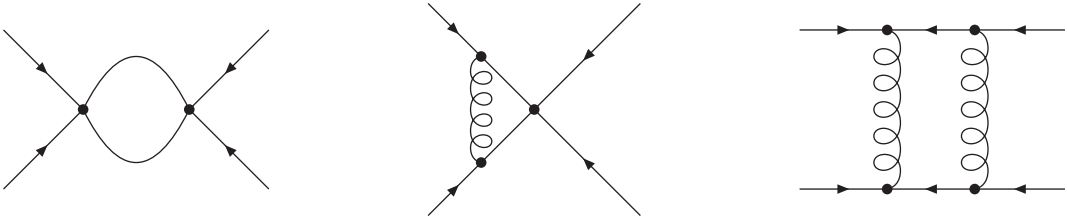
\end{picture}\\ {\sl fig.2}
\end{center}

\begin{center}
\begin{picture}(250,100)(0,0)

\ArrowLine(0,80)(30,50)
\ArrowLine(0,20)(30,50)
\ArrowLine(100,80)(70,50)
\ArrowLine(100,20)(70,50)
\Vertex(30,50){2}
\Vertex(70,50){2}
\Gluon(70,50)(30,50){5}{4}
\Gluon(15,65)(15,35){3}{3}

\ArrowLine(170,100)(200,70)
\ArrowLine(230,100)(200,70)
\ArrowLine(170,0)(200,30)
\ArrowLine(230,0)(200,30)
\Vertex(200,70){2}
\Vertex(200,30){2}
\Gluon(200,70)(200,30){5}{4}
\Gluon(215,85)(185,85){3}{3}
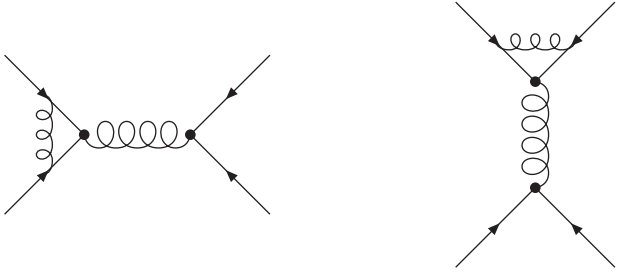
\end{picture}\\ {\sl fig.3}
\end{center}
Here the solid lines represent $\pi^\pm$ fields, the dashed lines
represent the $\theta$ scalar field, whilst the coiled lines are the
$\lambda_{\rho\sigma}$ antisymmetric tensor fields.

\end{document}